\documentclass[aps,prev,preprintnumbers,floatfix,nofootinbib,twocolumn]{revtex4-1}
\pdfoutput=1

\usepackage{mathrsfs, amsmath, amsthm, amssymb, color, epstopdf, verbatim, hyperref, enumerate,graphicx,hyperref}

\begin{document}

\title{Natural explanation for 21cm absorption signals via axion-induced cooling}
\author{Nick Houston$\,{}^a$}
\email{nhouston@itp.ac.cn}
\author{Chuang Li$\,{}^{a,b}$}
\email{lichuang@itp.ac.cn}
\author{Tianjun Li$\,{}^{a,b}$}
\email{tli@itp.ac.cn}
\author{Qiaoli Yang$\,{}^{c}$}
\email{qiaoli\_yang@hotmail.com}
\author{Xin Zhang$\,{}^{d,e}$}
\email{zhangxin@nao.cas.cn}
\affiliation{${}^a$ CAS Key Laboratory of Theoretical Physics, Institute of Theoretical Physics, Chinese Academy of Sciences, Beijing 100190, China\\
				${}^b$ School of Physical Sciences, University of Chinese Academy of Sciences, No. 19A Yuquan Road, Beijing 100049, China\\
				${}^c$ Siyuan Laboratory, Physics Department, Jinan University, Guangzhou 510632, China\\
				${}^d$ Key Laboratory of Computational Astrophysics, National Astronomical Observatories, Chinese Academy of Sciences, 20A Datun Road, Chaoyang District, Beijing, 100012, China\\
				${}^e$ School of Astronomy and Space Science, University of Chinese Academy of Sciences}

\begin{abstract}
The EDGES Collaboration has reported an anomalously strong 21cm absorption feature corresponding to the era of first star formation, which may indirectly betray the influence of dark matter during this epoch. 
We demonstrate that, by virtue of the ability to mediate cooling processes whilst in the condensed phase, a small amount of axion dark matter can explain these observations within the context of standard models of axions and axion-like-particles. 
The EDGES best-fit result favours an axion-like-particles mass in the (10, 450) meV range, which can be compressed for the QCD axion to (100, 450) meV in the absence of fine tuning.
Future experiments and large scale surveys, particularly the International Axion Observatory (IAXO) and EUCLID, should have the capability to directly test this scenario.

\end{abstract}

\maketitle

\textbf{Introduction.}
After recombination, between the thermal decoupling of baryons and the CMB and the era of first star formation, the Universe entered a prolonged period of cooling known as the dark ages.
Intriguingly, this epoch is both largely untested by observations, and in the standard $\Lambda$CDM cosmology, relatively predictable and easily understood.
As such, it can serve as a precise probe of physics outside of the $\Lambda$CDM paradigm.

One key observable is related to the absorption of 21cm light at that time, arising from the neutral hydrogen present filtering background radiation, and thereby imprinting a characteristic spectral distortion on wavelengths close to atomic transitions.
This feature redshifts to the 80 MHz range today, and has been recently observed by the Experiment to Detect the Global Epoch of reionisation Signature (EDGES) Collaboration. 

Their result is an anomalously strong 21cm absorption feature from $z\in (20,15)$, corresponding to the era of first star formation \cite{Bowman:2018yin}.
The amplitude of this signal is
\begin{equation}
	T_{21}\simeq 35 \mathrm{mK}\left(1-\frac{T_{\gamma}}{T_s}\right)\sqrt{\frac{1+z}{18}}
	\simeq -0.5^{+0.2}_{-0.5}\,\mathrm{K}\,,
	\label{T21}
\end{equation}
where $T_\gamma$ is the CMB temperature, $T_s$ the singlet/triplet spin temperature of the hydrogen gas present at that time, and the uncertainties quoted are at 99\% confidence level.
Once stellar emission of UV radiation begins at $z\sim 20$ we expect that $T_{\gamma}>>T_s\gtrsim T_{\mathrm{gas}}$, due to the decoupling of the CMB and hydrogen gas at $z \sim 200$, and the coupling of the spin temperature to the kinetic gas temperature.
In the standard $\Lambda$CDM scenario $T_{\gamma}|_{z\sim17}\simeq 49$ K and $T_{\mathrm{gas}}|_{z\sim17}\simeq 6.8$ K, so we expect $T_{21}\gtrsim-0.2$ K.
The resulting significance of this deviation from the $\Lambda$CDM prediction is estimated to be $3.8\,\sigma$.

One approach to resolving this discrepancy relies upon interactions with cold dark matter (CDM) to lower the gas temperature.
However, as demonstrated in Ref.~\cite{Barkana:2018lgd}, the interaction cross section required to achieve this is prohibitive for models of dark matter.
Consistency with other experimental and observational constraints ultimately limits models capable of explaining the EDGES observation to being comprised of just $0.3-2$\% millicharged dark matter, with masses and millicharges in the $(10, 80)$ MeV and $(10^{-4},10^{-6})$ ranges, respectively \cite{Munoz:2018pzp, Berlin:2018sjs,Barkana:2018qrx}
A number of other approaches have also been explored, including adding additional dark sector interactions, modifying the thermal history, and injecting additional soft photons during that epoch \cite{Fraser:2018acy, Costa:2018aoy,Li:2018kzs,Hill:2018lfx, Falkowski:2018qdj}.
Several axion-theoretic explanations have also been recently proposed \cite{Lambiase:2018lhs,Lawson:2018qkc,Moroi:2018vci}, but we emphasise for clarity that our approach differs in many essential respects from these.

More specifically, in the following we propose a dark-matter theoretic approach, which relies upon the speculated ability of axion dark matter to form a Bose Einstein Condensate (BEC) \cite{Sikivie:2009qn,Erken:2011dz}.
Whilst behaving in many respects as ordinary CDM, a particularly interesting aspect of this phenomenon exists in the ability of this condensed state to induce transitions between momentum states of coupled particle species and thereby mediate cooling processes.
This scenario was originally invoked in Ref.~\cite{Erken:2011vv} to lower the photon temperature in the era of Big Bang Nucleosynthesis (BBN), in order to adjust the baryon-to-photon ratio and thus ease the discrepancy between the observed and predicted primordial ${}^7$Li abundance.

As we will see in the following, by analogously lowering the hydrogen temperature prior to the cosmic dawn this mechanism can explain the EDGES observations in the context of axion and axion-like-particle (ALP) models.
The implied parameter range is close to existing experimental limits, and so could be tested at the next generation of axion experiments and via large scale surveys, particularly IAXO and EUCLID, respectively \cite{Armengaud:2014gea, Laureijs:2011gra}.

\textbf{Axion dark matter condensation.}
The underlying conditions for BEC formation are that a system comprise a large number of identical bosons, conserved in number, which are sufficiently degenerate and in thermal equilibrium \cite{Chakrabarty:2017fkd}.
As such, the formation of a BEC of CDM axions seems a reasonable possibility.

Nonetheless, there has been some controversy in the literature around this and the value of the resulting correlation length \cite{Saikawa:2012uk,Davidson:2013aba,Davidson:2014hfa,Guth:2014hsa}. 
In particular it was ultimately concluded in \cite{Guth:2014hsa} that although a Bose-Einstein condensate can form, the claim of long-range correlation in the case of attractive interactions is unjustified.
It has however been argued more recently that these findings may be overly reliant on the criterion of homogeneity, whilst a BEC can be inhomogeneous and nonetheless correlated over its whole extent, which can be arbitrarily large \cite{Chakrabarty:2017fkd}.
Addressing these points is somewhat beyond the scope of this paper, and so we instead proceed under the assumption that the BEC cooling mechanism functions as advertised in \cite{Sikivie:2009qn}.

It is nonetheless key to note that for generic ALPs with repulsive self-interactions, the thermalisation rate is
\begin{equation}
	\Gamma_a/H\sim4\pi \lambda n_a m_a^2/H\,,
	\label{ALP thermalisation}
\end{equation}
where $\lambda$, $n_a$ and $m_a$ are respectively the quartic coupling, and the cold axion number density and mass.
Since this increases with time, long range order will eventually be established and condensation can be reasonably and uncontroversially expected \cite{Guth:2014hsa}.

For the QCD axion, which provides both a compelling solution to the strong CP problem and a particularly attractive target for beyond the Standard Model physics searches \cite{Peccei:1977hh, Weinberg:1977ma, Wilczek:1977pj, Kim:1979if, Shifman:1979if, Dine:1981rt, Zhitnitsky:1980tq}, all available interactions are however attractive. 
The thermalisation rate due to gravitational interactions is given by
\begin{equation}
	\Gamma_a/H\sim4\pi Gn_am_a^2l_a^2/H\,,
\end{equation}
where $G$ is Newton's constant, and $l_a$ is  the correlation length \cite{Sikivie:2009qn}.
This scales as $t/a$, where $a$ is the scale factor, and so by the logic of \cite{Chakrabarty:2017fkd} can also be relied upon to ensure long-range order and the condensed phase persists.

Once formed, the large-scale gravitational field of the condensate can reduce the momenta of particle species, with the cooling effects beginning once the characteristic relaxation timescale $\Gamma$ exceeds the Hubble rate, so that
\begin{equation}
	\Gamma/H\sim4\pi Gm_an_al_a\omega/\Delta pH\gtrsim 1\,,
\end{equation}
where $\omega$ and $\Delta p$ are the energy and momentum dispersion of the particle species in question.
 
This phenomenon offers the possibility to then explain the anomalous EDGES result, with condensed axion dark matter cooling the primordial hydrogen after it decouples from the CMB at $z\sim 200$.
This latter point is essential, as if axion cooling begins whilst the CMB and hydrogen remain in thermal equilibrium, the effect on \eqref{T21} will be negligible.
Of course the onset of cooling must also be prior to the cosmic dawn, and the effect in total must give the correct EDGES absorption magnitude.
As we will see in the following, and perhaps surprisingly, these various requirements can be simultaneously accommodated by an ALP which may or may not also function as the QCD axion.
In practice the EDGES observation uniquely selects a small range for $m_a$, which is compatible with present-day axion phenomenology and can conceivably be explored at the next generation of axion experiments.

\textbf{Condensate-induced hydrogen cooling.}
Using the formulae of the previous section, our starting point is the baryon cooling rate at the time of matter-radiation equality,
\begin{equation}
	\frac{\Gamma_H}{H}\bigg |_{t_{eq}}\sim\sqrt{\frac{3m_H}{16 T_{eq}}}\frac{\Omega_ah^2}{\Omega_{DM}h^2}\,,
	\label{hydrogen cooling rate}
\end{equation}
where $\Omega_ah^2/\Omega_{DM}h^2$ is the fraction of the cooling-induced ALP density over the dark matter relic density, where we have used the Friedmann equation at this time to identify $3H^2\simeq16\pi G\rho_{DM}$, neglecting the contributions of visible matter and dark energy, and, assuming that we are in the condensed phase, identified $l_a\sim 1/H$.
By virtue of the Maxwell-Boltzmann distribution $\Delta p\simeq\sqrt{3m_HT_H}$, and at this temperature we can identify $\omega\sim m_H$.

As $m_H>>T_{eq}$ we evidently need a small $\left(\Omega_a/\Omega_{DM}\right)$ ratio to ensure cooling only begins when $z \in(200, 20)$.
To be more precise we can note that since $a\propto t^{2/3}$ during matter domination, $\Gamma_H/H\propto 1/\sqrt{T}$.
This then implies that after matter-radiation equality, 
\begin{equation}
	\frac{\Gamma_H}{H}=\frac{\Gamma_H}{H}\bigg |_{t_{eq}}\left(\frac{T_{eq}}{T_H}\right)^{1/2}\,.
\end{equation} 
Since $T_{eq}\sim 0.75$ eV $\simeq 8.7\times 10^3$ K, and we require axion-induced cooling to occur between $T_H^{z=200}\sim 475$ K and $T_H^{z=20}\sim 10$ K, we can first establish that we require 
\begin{equation}
	\Omega_a h^2/\Omega_{DM}h^2\in(0.22,1.5)\times10^{-5}\,.
	\label{CDM relic density}
\end{equation}

It is important to note that once the BEC forms, we will have two distinct populations of cold axions; those that are in the condensed state, and a remnant thermal population.
Hydrogen can in principle interact with both, however there exists a key distinction; scattering from the cold thermal axions will simply raise their temperature, whilst scattering from the condensed axions will typically liberate them from the BEC, given the energies involved, and into the thermal population. 
However, in Ref.~\cite{Davidson:2014hfa} the rate at which the BEC occupation number can change by scattering with external particles is calculated, finding that the latter number-changing process should be vanishingly rare.

Energy conservation then dictates that 
\begin{equation}
	\rho_H\left(T_i\right)\simeq\rho_H\left(T_f\right)+\rho_a\left(T_f\right)\,.
	\label{energy conservation}
\end{equation}
since the energy lost from the hydrogen must be transferred to the thermal axions
\footnote{For our parameter range of interest, photon cooling can be neglected.
As $n_H$ remains constant during axion cooling, to explain \eqref{T21} we require $\rho_H(T_i)/\rho_H(T_f)\sim \mathcal{O}(1)$, implying $\rho_H$ and $\rho _a$ must be of the same order.
Since $\rho_\gamma>> \rho_H$, from the known baryon-photon ratio, the resulting $\rho _a$ is too small to affect $\rho_\gamma$ and hence $T_{\gamma}$.
Thermal axion heating by photons is also strongly suppressed, as there is no large $\sqrt{m_H/T}$ factor in the corresponding equivalent of \eqref{hydrogen cooling rate}.
We also note the principal constraint in the axion-induced cooling ${}^7$Li scenario was a large resulting $N_{\mathrm{eff}}$ at recombination.
For us this is not a cause for concern as we are operating at a much later epoch, and the thermal axions excited will be non-relativistic.}.
In the case of cold hydrogen gas $\rho_H\simeq n_H(m_H+3T/2)$ to lowest order, where $n_H$ is the relic abundance.
Since hydrogen comprises the majority of baryonic matter at this epoch we can use the baryon-to-photon ratio to estimate $n_H\simeq 6\times 10^{-10}\,n_\gamma$, where $n_\gamma=2\zeta(3)T_\gamma^3/\pi^2$ is the photon number density. 
Inserting a Maxwell-Boltzmann distribution for the thermal axions we have
\begin{equation}
	\rho_a=\frac{T^4}{2\pi^2}\int_{0}^\infty\frac{\xi^2\sqrt{\xi^2+(m_a/T)^2}}{\exp\left(\sqrt{\xi^2+(m_a/T)^2}\right)- 1}d\xi\,,
\end{equation}
and we can solve \eqref{energy conservation} numerically for the cooling ratio $T_i/T_f$.

Assuming for simplicity that the change in $z$ is negligible during the cooling process, we have
\begin{equation}
	T_H^{z=17}\simeq T_H^{z=200}\left(\frac{z_{c}+1}{200+1}\right)^2\left(\frac{T_f}{T_i}\right)\left(\frac{17+1}{z_{c}+1}\right)^2\,,
\end{equation}
where $z_{c}$ is the redshift at which cooling begins.
Since dependence on this quantity cancels, we find
\begin{equation}
	T_{21}=35 \,\mathrm{mK}\left(1-\frac{T_i}{T_f}\frac{T_{\gamma}}{T_H}\right)\sqrt{\frac{1+z}{18}}\,,
	\label{Cooled T21}
\end{equation}
where $T_\gamma$ and $T_H$ take their usual $\Lambda$CDM values.
In practice additional care is needed since basic redshift relations do not accurately capture the evolution of $T_H$ in this region, so we use RECFAST to compute $T_H$ and $T_\gamma$ \cite{Seager:1999bc}. 
However, the resulting dependence in \eqref{Cooled T21} is nonetheless correct, and so we can use \eqref{energy conservation} to find the resulting 21cm absorption feature.
This is given in Fig.~1, where we see the EDGES best-fit value favours an ALP with mass $m_a\in\left(10,450\right)$ meV. 

\begin{figure}[h!]
	\centering
	\includegraphics[width=\linewidth]{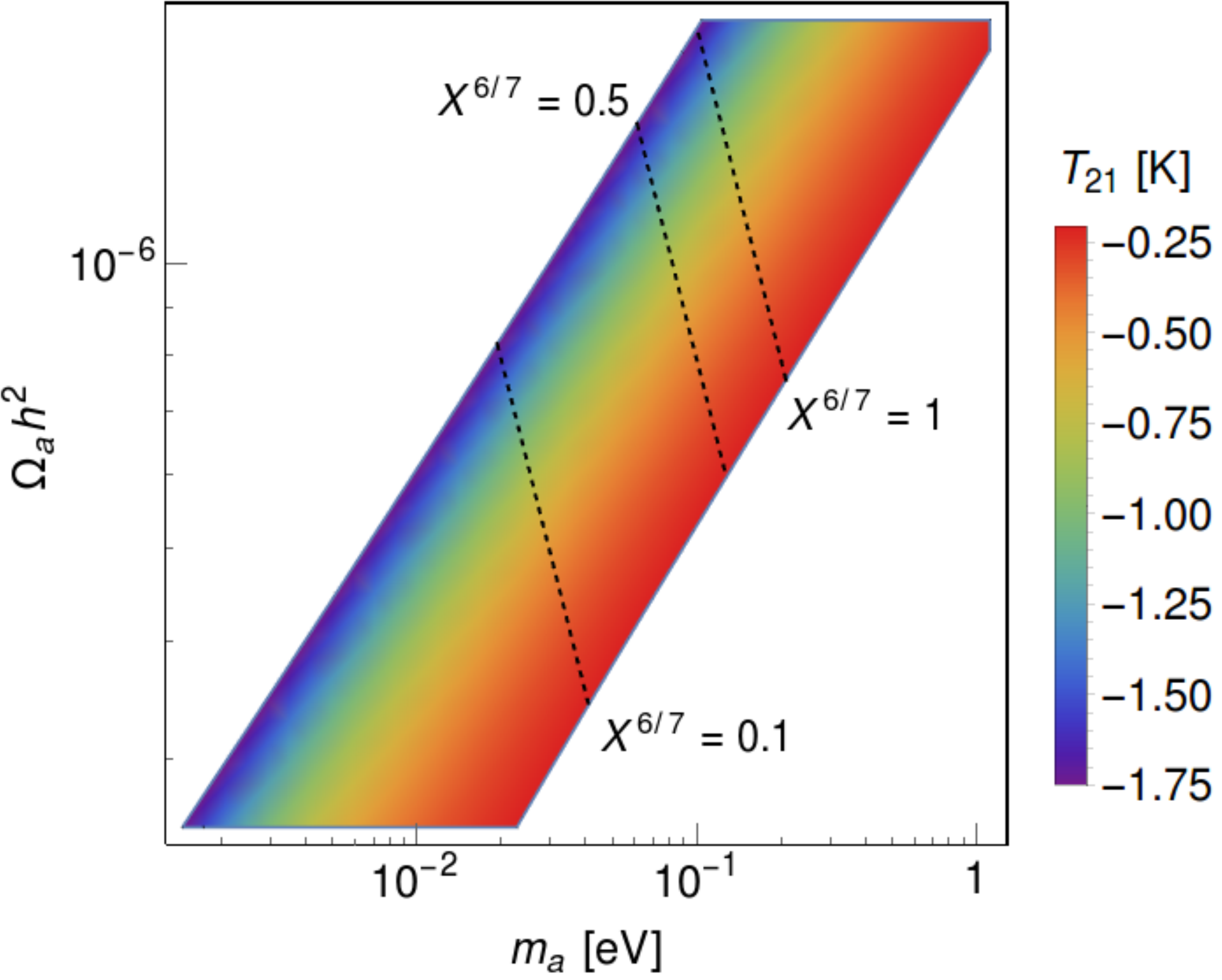}
  	\caption{The ALP $(m_a,\Omega_ah^2)$ space satisfying \eqref{CDM relic density}, colour-coded with the resulting 21cm brightness temperature at $z=17$. 
	Comparison with the best-fit EDGES result then suggests a $m_a\in\left(10,450\right)$ meV range of compatibility.
	Since the QCD axion fixes the relationship between these quantities in terms of the dark matter density parameter $X$ appearing in \eqref{QCD axion relic density}, we overlay lines of fixed $X$ to show the dependence on this quantity.}
\end{figure}

Since in the generic ALP case the relation between $\Omega_ah^2$ and $m_a$ is unfixed, we cannot directly connect them to coupling constraints and thus standard axion phenomenology.
However, for the QCD axion the corresponding $f_a$ is given via 
\begin{equation}
	\Omega_a h^2=0.15 X \left(\frac{f_a}{10^{12}\,\mathrm{GeV}}\right)^{7/6}\,,
	\label{QCD axion relic density}
\end{equation}
where for Peccei-Quinn (PQ) symmetry breaking before inflation, $X\sim \sin^2\theta_{mis}/2$, whilst for PQ symmetry breaking after inflation $X\in\left(2,10\right)$ depending on the relative contributions of topological defect decays and vacuum misalignment \cite{Erken:2011dz}.
This yields $f_a\in\left(1.2,6.1\right)\times X^{-6/7}\times 10^7$ GeV, which is in turn related through chiral perturbation theory to $m_a$ via
\begin{equation}
	m_a\simeq 6\,\mathrm{eV}\left(\frac{10^6\,\mathrm{GeV}}{f_a}\right)\,,
	\label{axion mass range}
\end{equation}
yielding $m_a\in\left(0.1,0.5\right) \times X^{6/7}$ eV.

Note however that $m_a$ is not freely varied in this case; each value is associated to a specific $\Omega_ah^2$, and thus the specific $z$ and $T_H$ at which cooling begins. 
Taking care to accommodate this, we find a one-to-one mapping between $m_a$ and $T_{21}$.
We also note for clarity that in this mass range we can expect both hot and cold axion dark matter, due, for example, to thermal production and vacuum misalignment respectively.

This being the case, we also represent the QCD axion in Fig.~1 via lines of constant $X^{6/7}$.
Since $X\in(2,10)$ for PQ symmetry breaking after inflation, the minimum value for this quantity is realised for pre-inflationary symmetry breaking.
In this case we have $X^{6/7}\sim 0.5$ in the absence of fine-tuning, assuming the initial misalignment angle is randomly drawn from a uniform distribution on $[-\pi,\pi]$, giving $\langle\theta_{\mathrm{mis}}^2\rangle=\pi^2/3$.

Varying $X$ we find a preferred natural range of $m_a\in(100,450)$ meV for the QCD axion by virtue of the EDGES best-fit result, where each value gives $T_{21}\simeq -0.5$ K at $z\sim17$ by solution of \eqref{energy conservation}. 
Fixing $X=1$ as a benchmark case we find $T_{21}\in( -1.75,-0.21)$ K at $z\sim17$, where we remind the reader that $T_{21}\simeq-0.21$ K is the standard $\Lambda$CDM result, which we reach in the limit of this mechanism being inoperative.
Working backwards, the 99\% confidence limits presented in \eqref{T21} then in this case imply the range $m_a\in(120,180)$ meV, with the best fit value corresponding to $m_a\simeq 150$ meV.

\textbf{QCD axion constraints.} 
Since the ALP case does not immediately translate to ordinary axion constraints, we can specialise to the QCD axion to gain some phenomenological insight and delineate the parameter values implied by the EDGES observation in this scenario, along with the various experimental and observational constraints which may apply.
In Fig.~2 we reproduce constraints on the axion parameter space in our region of interest from \cite{Graham:2015ouw} colour coded with the resultant value of $T_{21}$ at $z=17$ for the benchmark case of $X=1$. 
As can be seen, the EDGES observations can be straightforwardly accommodated within the ordinary QCD axion band.
Furthermore much of the resulting preferred parameter space will be covered by the IAXO experiment, allowing the possibility of a direct confirmation of these findings.
\begin{figure}[h!]
	\centering
	\includegraphics[width=\linewidth]{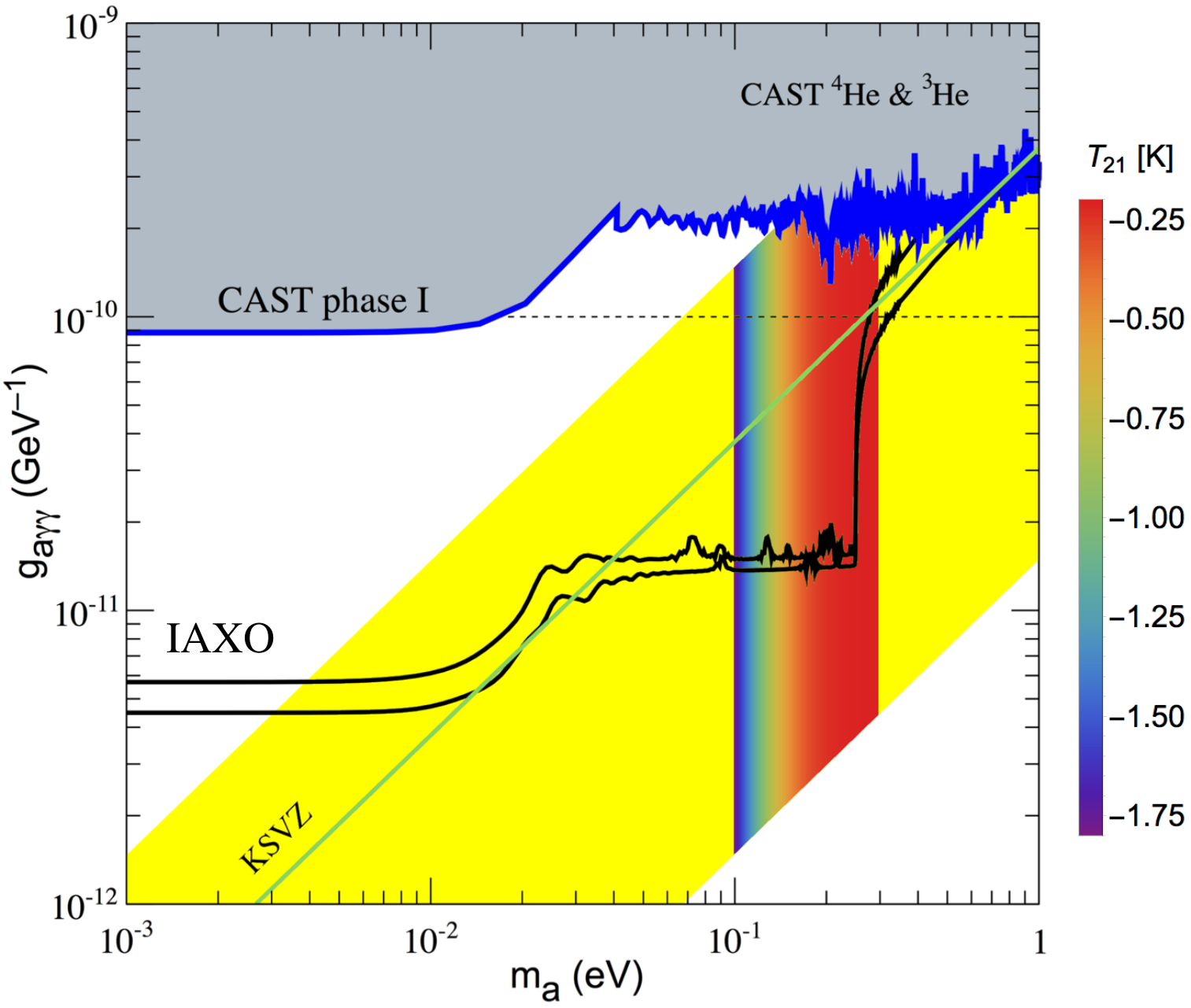}
  	\caption{The portion of the axion parameter space relevant for our purposes, reproduced from \cite{Graham:2015ouw}, with the 21cm brightness temperature at $z\sim17$ overlaid from axion-induced cooling processes in the benchmark case of $X=1$.
	The yellow band denotes QCD axion models with varying electromagnetic/colour anomaly coefficients, whilst the black curves indicate possible sensitivities for the proposed IAXO experiment. 
	The best fit axion mass value preferred by the EDGES observations in this case is 150 meV, whilst beyond $\sim 300$ meV the ordinary $\Lambda$CDM result for $T_{21}$ prevails.
	Variations in $X$ enlarge the preferred range to $m_a\in(100,450)$ meV in the absence of fine-tuning, effectively shifting the colour-coded region within the QCD axion band.}
\end{figure}

We can also note from Ref.~\cite{Archidiacono:2013cha} that although our mass range of interest evades hot dark matter constraints at present, future large scale surveys such as the EUCLID mission are in conjunction with Planck CMB data projected to probe $m_a\gtrsim150$ meV for the QCD axion at high significance, allowing this scenario to be definitively tested in the near future \cite{Archidiacono:2015mda}.

It is of course important to note that the full possible mass range favoured by these results is for DFSZ type axions strongly disfavoured due to stellar energy-loss arguments \cite{Zhitnitsky:1980tq,Dine:1981rt, Graham:2015ouw}.
As such we are implicitly considering KSVZ type models \cite{Shifman:1979if,Kim:1979if}, although the ratio $E/N$ of the electromagnetic to colour anomaly is however allowed to vary within the usual range to accommodate variant models of the QCD axion \cite{DiLuzio:2016sbl, DiLuzio:2017pfr}.

Strictly speaking even then there is tension between our preferred mass range and the observed burst duration of SN1987A, which favours $f_a\gtrsim 4\times 10^8$ GeV for standard QCD axions \cite{Raffelt:1996wa}. 
This arises from an inference of the supernova cooling timescale, and thus energy loss to axions, from the time interval between the first and last neutrino observation.
However, given that these limits are derived from a single observation, and not to mention our limited knowledge available about axion emission in this extreme environment (the resulting exclusion being `fraught with uncertainties' in the words of Ref.~\cite{Raffelt:1996wa}), we can follow the example of others (e.g. Ref.~\cite{Archidiacono:2013cha}) and exercise a measure of caution in applying this constraint.

So-called `astrophobic' axion models are also of note here, where $\mathcal{O}$(100) meV axion masses are allowed at the cost of introducing some flavour-violating couplings \cite{DiLuzio:2017ogq,Hindmarsh:1997ac}.
Furthermore, we can also recapitulate at this point that ultimately the axion cooling mechanism leveraged here is gravitationally mediated, and so could be achieved with no Standard Model couplings whatsoever, and thus no issues in this regard.
By extension, the use of the QCD axion is in this context non-essential, and our primary results for generic axion-like-particles can still apply nonetheless.

\textbf{Discussion and conclusions.}
The EDGES collaboration have recently presented an anomalously strong 21cm absorption profile, which may be the result of dark matter interactions around the time of the cosmic dawn.
Despite a flurry of interest there is as of yet no clear consensus on the provenance of this effect, and indeed whether it is a signature of dark matter at all, however these results nonetheless provide an exciting first window into a previously unexplored epoch.
 
We have in this letter explored the potential of condensed-phase axion dark matter, previously employed in the service of photon cooling, to explain these anomalous observations via reduction of the hydrogen spin temperature during this epoch.
By fixing the axion CDM relic density so that cooling begins within the appropriate epoch, we find that the resulting cooling effects are both capable of explaining the EDGES observations and compatible with present day axion phenomenology.

More specifically, we find that the EDGES best-fit result of $T_{21}\simeq-0.5$ K and the requirement that hydrogen cooling occur when $z\in (200,20)$ are consistent with the cooling induced by an axion-like-particle of mass $m_a\in\left(10,450\right)$ meV.
Specialising further to the QCD axion case, we find a preferred range $m_a\in(100,450)$ meV, in the absence of fine-tuning.

Furthermore, future experiments and large scale surveys such as IAXO and EUCLID should have the capability to directly probe the relevant parameter region and thereby test this scenario.
Indeed, as a dedicated direct-detection experiment sensitive in this mass range IAXO offers particular promise with regards to this scenario.
That said, as the underlying cooling mechanism relies only upon gravitational couplings it is not limited strictly to the context of models of the QCD axion, and so can also be arranged to occur in the primary scenario of axion-like-particles with no Standard Model couplings whatsoever, which could then evade these bounds.

We also note Ref.~\cite{Sikivie:2018tml}, which appeared shortly after this letter appeared online and deals with exactly the same scenario of axion BEC-induced cooling and 21cm cosmology.
A key point raised therein, which we have not previously addressed, is that this mechanism may have a damping effect on Baryon Acoustic Oscillations (BAO). 
Although it is ultimately argued there that the net effect on BAO should be consistent with observations, it may be worthwhile to more deeply explore the consequences of this scenario for this and other cosmological observables.

\begin{acknowledgments}
This research was supported by a CAS President's International Fellowship, the Projects 11475238, 11647601, 11875062 and 11875148 supported by the National Natural Science Foundation of China, and by the Key Research Program of Frontier Science, CAS.
We also thank our anonymous referees for their very helpful comments and suggestions.
\end{acknowledgments}

\end{document}